**Geographical Variation in Project Cost Performance: The Netherlands versus Worldwide**

By

Chantal C. Cantarelli, Bent Flyvbjerg, and Søren L. Buhl

## Abstract

Cost overruns in transport infrastructure projects know no geographical limits; overruns are a global phenomenon. Nevertheless, the size of cost overruns varies with location. In the Netherlands, cost overruns appear to be smaller compared to the rest of the world. This paper tests whether Dutch projects perform significantly better in terms of cost overruns than other geographical areas. It is concluded that for road and tunnel projects, the Netherlands performs similarly to the rest of the world. For rail projects, Dutch projects perform considerably better, with projects having significantly lower percentage cost overruns in real terms (11%) compared to projects in other North West European countries (27%) and in other geographical areas (44%). Bridge projects also have considerably smaller cost overruns – 7% in the Netherlands compared with 45% in other NW European countries and 27% in other geographical areas. In explaining cost overruns, geography should therefore clearly be taken into consideration.

*Keywords:* cost overrun, geography, Netherlands, worldwide, rail projects

## Introduction

Whether it is in Europe, America, Australasia or elsewhere, all over the world, examples of large cost overruns in transport infrastructure projects can be found. One of the most famous "project disasters" in this respect is the Channel tunnel. This undersea rail tunnel linking the United Kingdom and France is the longest in its kind in Europe with a length of about 50 kilometres. Construction costs increased from £2600 million to £4650 million (1985 prices) – 80 per cent higher than forecasted (Flyvbjerg et al., 2003a). In Boston, in the United States, a much disputed project is the Central Artery/Tunnel project, a large and complex underground highway project. The project, also known as the "Big Dig" or "Big Dug" due to persistent tunnel leaks, had overrun its costs by US $ 11 million or 275 per cent when it opened (Flyvbjerg 2007, EPB). The Mass Rapid Transit, the underground metro system in Bangkok which is about 20 kilometres long and was 67% over budget when completed can be seen as yet another transport infrastructure failure. Cost overruns appear to be a global phenomenon, existing across 20 nations on five continents (Flyvbjerg et al., 2003b).

Few studies comparing actual and estimated costs have taken the geographical location into account. To the authors' knowledge, the study by Flyvbjerg et al. (2003a, 2003b) is the only study that tests whether cost overruns vary with geographical location. This study covers 258 projects (roads, rail, tunnels and bridges) and distinguishes three geographical areas: Europe, North America and "other geographical areas (a group of 10 developing nations plus Japan)." They found statistically significant differences in average cost overruns between these areas with average overruns typically being smaller for Europe (26%) and North America (24%) compared to other areas (65%). Thus, geography matters for cost performance. However, Flyvbjerg et al. (2003b) argue that the highly significant differences in cost escalation depending on geographical location are due to the large average cost overrun of the projects in the "other geographical areas", "with their poor track record of cost escalation for rail, averaging 64.6% (Flyvbjerg et al. 2003b)."

Although cost overruns seem to be a global phenomenon, they also appear to vary with geographical location. This is supported by a recent study by Cantarelli et al. (forthcoming-a and -b) which shows that the cost performance of transport infrastructure projects in the Netherlands is actually quite different from the worldwide findings. The main differences are:

- Cost overruns are not predominant but are as common as cost underruns.
- The average cost overrun is considerably smaller than the worldwide average.
- Rail projects have the smallest overruns whereas worldwide rail projects have the largest average cost overrun.

The purpose of this paper is to establish the extent to which cost performance in Dutch projects differs with geographical location. This geographical variation will be examined within the context of the Netherlands and within a worldwide context. In addition, possible explanations for the findings are provided.

In order to address these objectives, the sample from the Dutch study is used to analyse in a statistically valid and reliable manner the extent to which the cost performance in the Netherlands differs from the rest of the world and from specific geographical locations. For the international data, an enlarged version of the original database (Flyvbjerg et al., 2003b), now including 806 projects instead of 258 projects, is used. This immediately raises the question of whether the increase in the number of projects has any influence on the overall cost performance of the worldwide study. New countries have been included compared with the original database, and if geographical location plays a role, this could affect the overall performance. This will also be covered in this paper.

In sum, the following research questions are addressed: 1. To what extent do cost overruns within the Netherlands depend on geographical location? 2. What is the influence of the increase in the number of projects in the worldwide database on the cost performance? And 3. To what extent is the cost performance in the Netherlands different from that worldwide? The large worldwide database allows a distinction to be made between different regions of the world. For this paper, other North West (NW) European countries are considered in more detail.

The remaining part of this paper is as follows. In section 2, the data and methodology of the Dutch and international sample are described. Section 3 examines whether cost performance varies depending on geographical location within the Netherlands. Section 4 presents the cost performance of the 806 projects against the original cost performance based on 258 projects. The Dutch project performance is compared against that of the rest of the world in section 5 and that of other NW European countries and other geographical areas in section 6. Section 7 focuses on the most important results and provides explanations for the findings. Finally, section 8 presents the conclusions and discussion.

## Data and Methodology

This section presents the main characteristics of the Dutch and international samples. Both samples focus on land-based transport infrastructure projects, including the following four project types: roads, rail, tunnels and bridges.

### Dutch and International Data

*Characteristics of the Dutch Data*

After three years of data collection and refinement, a database was established consisting of 78 Dutch projects, which are distributed as follows over the four project types:

- Road: 37
- Rail: 26
- Tunnel: 8
- Bridge: 7

The Netherlands is divided into twelve provinces, but for the purpose of this research, 6 regions are distinguished[1]:

- North Netherlands (N NL): including provinces Friesland, Groningen, Drenthe
- East Netherlands (E NL): including provinces Gelderland, Overijssel
- Central Netherlands (C NL): including provinces Utrecht, Flevoland
- South Netherlands (S NL): including provinces Limburg, Noord-Brabant
- Noord-Holland (NH): province Noord-Holland
- Zuid-Holland (ZH): province Zuid-Holland

Table 1 provides an overview of the number of projects by geographical location and project type. Two rail projects i.e. the Betuweroute and HSL-South are implemented in more than one region and are marked as "crossing".

---

[1] Note that the database does not include any projects from the province Zeeland

**Table 1 Number of projects by geographical location and project type**

| Region | Project type | | | | Total | |
|---|---|---|---|---|---|---|
| | Road | Rail | Tunnel | Bridge | # | % |
| North Netherlands (N NL) | 7 | 2 | - | - | 9 | 12 |
| East Netherlands (E NL) | 3 | - | - | 1 | 4 | 5 |
| Central Netherlands (C NL) | 4 | 4 | 1 | 1 | 10 | 13 |
| South Netherlands (S NL) | 6 | 2 | 1 | - | 9 | 12 |
| Noord-Holland (NH) | 7 | 8 | 1 | 2 | 18 | 23 |
| Zuid-Holland (ZH) | 10 | 8 | 5 | 3 | 26 | 33 |
| Crossing | - | 2 | - | - | 2 | 3 |
| **Total** | **37** | **26** | **8** | **7** | **78** | **100** |

The largest share of projects (about 56%) has been implemented in the provinces Noord-Holland and Zuid-Holland. These are also the two most densely populated provinces with 1246 and 999 inhabitants per square kilometre respectively. These provinces contain three of the four largest cities in the Netherlands, being Amsterdam in the Noord-Holland province and Rotterdam and The Hague in the Zuid-Holland province. Together with the city Utrecht (C NL) and their surrounding areas they form the Randstad, which is the 6th largest metropolitan area (in terms of population size) in Europe.

*Characteristics of the International Data*

At the time of this research, the number of projects in the international database has about tripled compared to the 2003 study to reach a total number of 806 projects. It should be noted that the Dutch projects are included in this number for the purpose of comparing this enlarged database with the original database of 258 projects (section 4). However, when the cost performance of the Dutch projects is compared with the rest of the world (section 5 and 6), the Dutch projects are removed from the international database. Hence the 78 Dutch projects are compared with the 728 international projects. The 806 projects, which include the Dutch projects, are distributed as follows over the four project types:

- Road: 537
- Rail: 195
- Tunnel: 36
- Bridge: 38

The projects are located in 8 different regions, covering 17 countries plus two categories (S Eu and other developing countries) which comprises several unspecified countries[2]:

- North and West (NW) Europe (NW Eu): Netherlands, Switzerland, Germany, Denmark, France, Norway, Sweden, United Kingdom and Hungary
- South Europe (S Eu): specific countries in this region are unknown
- East Europe (E Eu): Slovenia
- North America (N Am): Canada and the US
- Latin America (L Am): Mexico
- Asia: Japan, South Korea and Thailand
- Africa: Zambia
- Other developing countries[3]

Table 2 gives an overview of the number of projects by geographical distribution and project type.

**Table 2 Characteristics of the international database**

| Region | Project type | | | | Total | |
|---|---|---|---|---|---|---|
| | Road | Rail | Tunnel | Bridge | # | % |
| NW Europe (NW Eu) | 315 | 90 | 32 | 22 | 459 | 57 |
| South Europe (S Eu) | 16 | 7 | - | - | 23 | 3 |
| East Europe (E Eu) | 37 | - | - | - | 37 | 5 |
| North America (N Am) | 24 | 65 | 3 | 16 | 108 | 13 |
| Latin America (L Am) | - | 1 | - | - | 1 | - |
| Asia | 138 | 20 | 1 | - | 159 | 20 |
| Africa | 7 | - | - | - | 7 | 1 |
| Other developing countries | - | 12 | - | - | 12 | 1 |
| **Total** | **537** | **195** | **36** | **38** | **806** | **100** |

The largest share of projects (about 57%) has been implemented in NW Europe. Since the Netherlands is geographically located in this area, the project performance of the Netherlands will be compared more specifically with this region. The other regions (being South Europe, East Europe, North America, Latin America, Asia, Africa and other developing countries) are pooled together into the group "other geographical areas."

---

**Methodology in the Dutch and International Study**

With the objective of determining whether the worldwide findings also apply for one specific country (in this case the Netherlands), the Dutch study followed from the outset as much as possible the same methodology as the worldwide research by Flyvbjerg et al. (2003a; 2003b). Projects in the international database were selected on the basis of data availability, that is, all the projects that were known and for which data were available for the development of construction costs were considered for inclusion in the sample. Since the methodology is the same, as with the original sample of 258 projects, the current sample of 806 projects is probably not representative of the population of transport infrastructure projects (see Flyvbjerg et al., 2003b). The sample is biased and the bias is conservative; the difference between the actual and estimated costs from the sample is likely to be lower than the difference in the project population. Despite this conservative bias, given the current state-of-the-art in this field of research, it is the best obtainable sample (see Flyvbjerg et al. (2003b) for an extensive description of the methodological considerations and implications of the international sample).

The intention of the Dutch study was to include all projects from a specified period in the database. However, due to the non-availability of key information and the incompleteness of information for some projects, the database does not cover all projects. However, in line with previous international research in this field that also included only projects for which information was available, the database is treated as a sample (see Cantarelli et al. (forthcoming-a) for a full description of the methodology in the Dutch study). Similar to the international study, the sample is probably biased and the bias is conservative. This is mainly caused by fixed link projects for which data was collected by means of interviews.

**Similarities and Differences between the Dutch and International Sample**

Both the Dutch and the worldwide sample include the same variables and the way in which the data for these variables are collected is also the same. The most important variables are the following:

- *Time of formal decision to build (ToD):* this is one specific point in the process when a decision was made to go ahead with the project, that is, the "go-decision"

- *Estimated opening year:* this is the expected year of opening at the ToD. If the estimated opening year is unavailable at the ToD, then the nearest available estimate of the opening year is used as a baseline.
- *Actual opening year:* year in which operations begin[4]
- *Estimated costs:* the costs at the ToD. When the costs are not available at the ToD, the nearest available reliable figure for estimated costs is used as a proxy.
- *Actual costs:* the costs at the actual opening year. If the actual costs are unknown at the time of project completion, the most reliable later figure for actual costs is used (i.e. from a year later than the opening year), if available. If unavailable, an earlier figure for actual costs could be used (i.e., from a year before the opening year), but only if 90% of the budget was spent at this time, i.e., the project was 90% complete in financial terms.

The way in which scope changes are handled can potentially have an impact on the difference in average cost overrun. Therefore in both the Dutch and the worldwide sample, scope changes are treated in the same way. Scope changes are included to the extent that the planned and implemented projects remain functionally identical. The project has to fulfil the same objective and serve the same market to be considered to have the same project function. If the project function remained the same over the years, the project was included in the research. If the project function at the ToD was different from the project function at the time of opening, an attempt was made to make the projects comparable. If this attempt failed, it was considered meaningless to compare the projects, and it was not included.

Besides the variables and scope changes, the overall way in which data is handled is the same in both the Dutch and the worldwide study. This makes it possible to compare the Dutch and worldwide data, which was almost impossible with previous studies on cost overruns (e.g. Auditor General of Sweden, 1994 (in Odeck 2004); Flyvbjerg et al., 2003a; Merewitz, 1973; Morris, 1990; Nijkamp and Ubbels, 1999; Odeck, 2004 and Pickrell, 1990,1992). The four main reasons for this are: i) the difference in use of nominal and real prices (Flyvbjerg, 2007), ii) different use of the time of the formal decision to build and actual opening year as a basis for the estimated and actual costs (Flyvbjerg et al., 2003b), iii) different sample size and iv) different geographical area. Both the Dutch and international sample present costs excluding VAT and correct for inflation using the appropriate geographical, sectoral and historical indices. Estimated and actual costs are based on the same base year (see above). Although the sample size of the Dutch projects is smaller compared to the international database it is still considered large enough to allow statistical analyses. This leaves only one main reason for differences found between both samples, namely geographical area, and this is exactly the subject under scrutiny in this paper. If cost overruns differ between both samples, this can be explained by geographical area. The main differences between both samples concern the *sources of data collection* and the *selection of large-scale projects*. In the worldwide sample, one of the core sources of information on the costs of projects is the National Audit Office. Instead, one of the main sources for data collection for the Dutch projects is the MIRT (*Meerjarenprogramma Infrastructuur, Ruimte en Transport*, translated as the Multi-years programme for infrastructure, spatial planning and transport)[5]. The different use of sources also resulted in a different approach to selecting projects.

---

[4] For projects that were based on the MIRT documentation, data on this actual opening year were unavailable and an assumption had to be applied (see Cantarelli et al., forthcoming -a)

## Geographical Variation in Cost Performance in the Netherlands

Table 3 provides an overview of the characteristics of the Dutch projects regarding the number of projects and average cost overruns by geographical region and project type.

**Table 3 Characteristics Dutch database**

| Region | Project type | | | | Total | | Cost overrun | |
|---|---|---|---|---|---|---|---|---|
| | Road | Rail | Tunnel | Bridge | # | % | Mean | SD |
| North Netherlands (N NL) | 7 | 2 | - | - | 9 | 12 | 11.6 | 35.9 |
| East Netherlands (E NL) | 3 | - | - | 1 | 4 | 5 | 9.5 | 24.9 |
| Central Netherlands (C NL) | 4 | 4 | 1 | 1 | 10 | 13 | 7.2 | 39.2 |
| South Netherlands (S NL) | 6 | 2 | 1 | - | 9 | 12 | 23.8 | 48.9 |
| Noord-Holland (NH) | 7 | 8 | 1 | 2 | 18 | 23 | 13.3 | 27.4 |
| Zuid-Holland (ZH) | 10 | 8 | 5 | 3 | 26 | 33 | 21.9 | 49.4 |
| Crossing | - | 2 | - | - | 2 | 3 | 29.0 | 36.8 |
| **Total** | **37** | **26** | **8** | **7** | **78** | **100** | **16.5** | **40.0** |
| Cost overrun   Mean | 18.8 | 10.6 | 34.9 | 6.6 | | | | |
| SD | 38.9 | 32.2 | 67.4 | 33.4 | | | | |

The average cost overrun is largest for the two projects that are cross-regional (28.9%, SD=36.5) followed by projects in the area South Netherlands (23.8%, SD=48.9). The geographical area with the smallest cost overrun is Central Netherlands with on average a cost overrun of 7.1% (SD=39.3). The average cost overrun is the largest for tunnel projects followed by road, rail and bridges.

---

[5] The translation of the MIRT in English is based on:
http://www.verkeerenwaterstaat.nl/english/topics/water/delta_programme/rules_and_framework_of_the_mirt (consulted 20-03-2010)

The dataset regarding region and type is quite unbalanced, that is, some combinations are included more often than others and some combinations are even not included at all. Still, a two-way ANOVA with sequential backwards elimination could be conducted. We started with the most complete model including the two individual factors and the interaction effect (base model) and tested this against the additive model, i.e. the model without the interaction effect (test model). It turned out that the test model could be accepted (F=1.345, p=0.229). Thus the difference in average cost overrun between regions is not more affected by the type of projects that are implemented in these regions than what could be ascribed to chance. The next step was to test whether one of the two models with only one the individual factors could be accepted (either the region factor or the type factor). It turned out that either of the simpler models could be accepted (F=0.793, p=0.502 for excluding project type and F=0.254, p=0.956 for excluding region). Finally, we excluded the factor with the highest p-value (region) and tested the effect of project type alone. Even that could be ascribed to chance (F=0.930, p=0.431). In the data there is a clear difference in average cost overrun between regions and between project types, but these differences could be ascribed to chance.

## Cost Overruns in 806 Projects Compared with Previous Data

This section determines whether and to what extent the cost performance of the projects in the new dataset comprising 806 projects differs from that of the projects in the original dataset of 258 projects. First we examine whether the representation of projects over geographical location and project type has changed with the increase of the database. This may give an indication of whether the cost performance has changed.

Table 4 presents the number of projects per region and project type for both databases. Note that the original 258 projects are also included in the enlarged international dataset of 806 projects. Further, in the original study based on 258 projects, fixed links were not broken down into tunnels and bridges, and hence we cannot present the number of projects per region for these project types separately either.

**Table 4 Number of projects per region and project type in the database with 806 and 258 projects**

| Region | Worldwide database (N=806) | | | | | Worldwide database (N=258) | | | |
|--------|------|------|--------|--------|-------|------|------|----------------|-------|
| | Road | Rail | Tunnel | Bridge | Total | Road | Rail | Fixed links | Total |
| NW EU | 315 | 90 | 32 | 22 | 459 | 143 | 23 | 15 | 181 |
| S EU | 16 | 7 | - | - | 23 | - | - | - | 0 |
| E EU | 37 | - | - | - | 37 | - | - | - | 0 |
| N Am | 24 | 65 | 3 | 16 | 108 | 24 | 19 | 18 | 61 |
| L Am | - | 1 | - | - | 1 | - | 1 | - | 1 |
| Asia | 138 | 20 | 1 | - | 159 | 0 | 3 | 0 | 3 |
| Africa | 7 | - | - | - | 7 | - | - | - | 0 |
| Other | - | 12 | - | - | 12 | - | 12 | - | 12 |
| **Total** | **537** | **195** | **36** | **38** | **806** | **167** | **58** | **33** | **258** |

The enlarged database of 806 projects differs in three aspects from the original database. First of all, projects from three new regions are included. These are projects in South Europe, East Europe and Africa. Secondly, Asian projects are better represented in the larger database. Thirdly, the number of projects for all three project types has been greatly increased. Despite a doubling in the number of fixed link projects, this increase remains the smallest compared to road and rail projects. Furthermore, the increase in fixed link projects did not incorporate any new geographical areas.

To conclude, the representation of projects regarding geographical location and project types has changed with the increase of the number of projects and a different cost performance could happen. Since no new fixed link projects from different regions were added, we do not expect a change for this project type.

Table 5 presents the number of projects, the mean cost overrun and the standard deviation for the worldwide samples with 806 projects and 258 projects.

**Table 5 Cost overruns broken down by project type for worldwide samples (N=806) and (N=258)**

| Project Type | Worldwide N=806 | | | Worldwide N=258 | | |
|--------------|------|------|------|------|------|------|
| | N | Mean | SD | N | Mean | SD |
| Road | 537 | 19.8 | 31.4 | 167 | 20.4 | 29.9 |
| Rail | 195 | 34.1 | 43.5 | 58 | 44.7 | 38.4 |
| Fixed Links | 74 | 32.8 | 58.2 | 33 | 33.8 | 62.4 |
| Bridges | 38 | 30.3 | 60.6 | n/a | n/a | n/a |
| Tunnels | 36 | 35.5 | 56.3 | n/a | n/a | n/a |
| **Total** | 806 | | | 258 | | |

In which n/a: not available (figures for bridges and tunnels in the original database of 258 were not available).

Table 5 shows that of the four project types, road projects have the smallest overrun of 20% followed by bridge projects with an overrun of 30%, rail projects with an overrun of 34% and tunnel projects with an overrun of 35%. Based on an F-test we conclude with overwhelming statistical significance that roads, rail, tunnels and bridges are different (F=8.293, p<0.001). Hence, project types should be treated separately when discussing cost overruns. However, considering the relatively small number of observations and similar cost performances, it could be argued that bridges and tunnels should be treated as one project type, fixed links. Based on a t-test, it is concluded that cost overruns do not significantly differ between bridges and tunnels (p=0.706). Hence, tunnels and bridges could be merged.

Furthermore, the average overruns come with large standard deviations, indicating that the data for individual projects are spread over a large range of values. Road projects have the smallest standard deviation indicating that on average road projects are nearer to the mean value of the overrun compared to rail, bridges and tunnels. A Bartlett test shows that the standard deviations of the different project types are different with very high statistical significance (p<0.001).

Considering the substantial and significant difference in mean cost overruns and standard deviations we have found, we conclude that project type matters and pooling the project types together is therefore not appropriate. In the analyses that follow, each type of project will therefore be considered separately.

Following these findings on cost overruns worldwide, let us compare these new results with the original results based on 258 projects. Looking at table 5 two figures immediately stand out; the considerably lower average cost overrun for rail projects and the hardly changed average overrun for road projects. From this we must conclude that geographical location has a larger influence on the average cost overrun for rail projects than for road projects. The decrease in average cost overrun for rail projects can be explained by the increase in the number of projects in Europe and North America, the areas with a better cost performance record. As expected, the cost performance of fixed link projects changed only slightly.

## Dutch Cost Performance versus the Rest of the World

Table 6 presents the number of projects, the mean cost overrun and the standard deviation for the 78 Dutch projects and for the 728 projects in the rest of the world.

**Table 6 Cost overruns broken down by project type: Netherlands versus rest of the world**

| Project Type | Netherlands N=78 | | | Rest of the world N=728 | | |
|---|---|---|---|---|---|---|
| | N | Mean | SD | N | Mean | SD |
| *Road* | 37 | 18.8 | 38.9 | 500 | 19.9 | 30.9 |
| *Rail* | 26 | 10.6 | 32.2 | 169 | 37.7 | 44.0 |
| *Fixed Links* | 15 | 21.7 | 54.4 | 59 | 35.7 | 59.2 |
| *Bridges* | 7 | 6.6 | 33.4 | 31 | 35.7 | 64.4 |
| *Tunnels* | 8 | 34.9 | 67.4 | 28 | 35.6 | 54.1 |
| ***Total*** | 78 | | | 726 | | |

Comparing the Netherlands with the rest of the world, the largest differences can be seen between the average cost overrun for rail and for bridges. Rail projects in the Netherlands have considerably smaller average cost overruns (11%) compared to the rail projects in the rest of the world (38%). Similarly, the average cost overrun for Dutch bridge projects, at 7%, is considerably smaller than the worldwide average of 36%. The difference in average is statistically significant for rail projects (p<0.001), but not for bridge projects (p=0.106). The non-significance for bridge projects is probably caused by the small number of projects. Still, considering the difference in average we must conclude that there is a very large and relevant difference from the other project types. For road and tunnel projects, the average cost overrun in the Netherlands is not significantly different from the other projects (p=0.875 and p=0.977 for roads and tunnels respectively).

# Cost Overruns in the Netherlands versus NW Europe and Other Geographical Areas

Section 2 already specified that the focus will be on NW European countries; the other regions are pooled together as "other geographical areas." First the cost performance in North West Europe is considered more closely.

## Cost Performance in North West Europe

Table 7 presents the number of projects, the mean cost overrun and the standard deviation for North West European countries broken down by project type.

**Table 7 Cost overrun for NW European countries by project type**

| Project Type | NW Europe | | |
| --- | --- | --- | --- |
| | N | Mean CO % | SD |
| *Road* | 315 | 20.9 | 30.2 |
| *Rail* | 90 | 22.3 | 34.9 |
| *Fixed links* | 54 | 31.5 | 48.6 |
| *Bridges* | 22 | 32.9 | 50.6 |
| *Tunnels* | 32 | 30.6 | 48.0 |
| *Total* | 459 | | |

Table 7 shows that again road projects have the best cost performance, closely followed by rail projects. Bridges and tunnels also perform rather similarly to each other. An F-test showed that the difference in average cost overrun between the project types for NW European countries is not statistically significant (F=1.533, p=0.205). This implies that, for this geographical area, project type does not matter for cost overruns and hence, projects could be pooled together. However, a Bartlett test shows that there is a highly significant (p<0.001) difference in the standard deviations of the project types. Since homogeneity of standard variances is a precondition for an F-test, and this has thus been violated, the different types of projects are considered separately in the following analyses. The most remarkable difference between the worldwide projects and NW European projects is that in NW Europe, the average overrun for rail is smaller.

**Cost Overruns in the Netherlands versus NW Europe and Other Geographical Areas**

These abovementioned figures for NW Europe include the Dutch projects, but if these projects are compared with other NW European countries and other geographical areas, a different picture emerges. Table 8 presents the number of projects, the mean cost overrun and the standard deviation for these three geographical areas broken down by project type.

**Table 8 Cost overruns in the Netherlands, other NW European countries and other geographical areas**

| Project Type | The Netherlands | | | Other NW European countries | | | Other geographical areas | | |
| --- | --- | --- | --- | --- | --- | --- | --- | --- | --- |
| | N | Mean | SD | N | Mean | SD | N | Mean | SD |
| *Road* | 37 | 18.9 | 38.9 | 278 | 21.2 | 28.9 | 222 | 18.2 | 33.1 |
| *Rail* | 26 | 10.6 | 32.2 | 64 | 27.1 | 35.0 | 105 | 44.2 | 47.6 |
| *Fixed links* | 15 | 21.7 | 54.4 | 39 | 35.3 | 46.4 | 20 | 36.4 | 80.0 |
| *Bridges* | 7 | 6.6 | 33.4 | 15 | 45.1 | 53.5 | 16 | 26.8 | 73.8 |
| *Tunnels* | 8 | 34.9 | 67.4 | 24 | 29.2 | 41.4 | 4 | 74.5 | 104.2 |
| *Total* | 78 | | | 381 | | | 347 | | |

Considering road projects, there does not seem to be a large difference in project performance between the different geographical regions with cost overruns ranging between 18% for other geographical areas, 19% for Dutch projects and 21% for other NW European countries. Based on a t-test (the Welch version was used because of problems with variance homogeneity), it was confirmed that the differences in average cost overrun between the Netherlands and other NW European countries is not significant (p=0.714). The difference in average cost overrun between the Netherlands and other geographical areas is not statistically significant either (p=0.934). However, for rail, bridges and tunnels, the average cost overruns do largely differ between the regions with differences in average cost overrun of about 30% for rail, approximately 20% for tunnels and even up to about 40% for bridges.

For rail projects, Dutch projects have significantly smaller cost overruns of 11% compared to the average overrun of 27% in other NW European countries (p=0.037) and an average of 44% in other geographical areas (p=0.001). Cost overruns for rail projects vary with geographical location.

Dutch projects again have the smallest average cost overrun for bridges. The difference with the other regions is quite large with a 7% overrun for Dutch projects, 45% for other NW European countries and 27% for other geographical areas. It seems as if the Netherlands clearly performs better than the rest of the world. However, the differences are not statistically significant (p=0.054 for the difference in average overrun with other NW European countries and p=0.376 for the difference with other geographical areas). The reason could be the small number of projects and/or the large standard deviations. As it is, the differences could be due to chance.

Lastly, considering the project performance of tunnel projects, NW European projects have the smallest average cost overrun of 29%, followed by Dutch projects with an average of 35% and projects in other geographical areas with an average cost overrun of 75%. Although the cost overrun in the other geographical areas seem much higher compared to the Dutch projects, the difference in average cost overrun is non-significant (p=0.525). The cost performance in the Netherlands does not significantly differ from that of other NW European countries either (p=0.827).

## Explanations

The analyses presented in the previous sections show some remarkable results. This section elaborates upon these results and provides explanations for these findings.

### Worldwide Cost Performance

There are two findings that are particularly remarkable in the study regarding the worldwide cost performance. First of all, the worldwide cost performance varies with project type; the cost overruns between road, rail, tunnels and bridges are significantly different with tunnel and bridge projects having on average the largest cost overrun. This has important policy implications. Cost estimates should be considered with care for all project types and in particular for fixed link projects. As to the reason why fixed link projects have the largest overruns, there are several plausible explanations. As a matter of fact, any or all of the different types of explanations - technical, political-economic or psychological explanations (Flyvbjerg et al. 2002, 2007), can apply. Technical explanations consider cost overruns to be the result of "forecasting errors" in technical terms e.g. imperfect forecasting techniques, inadequate data and lack of experience. The indivisibility argument, that is, projects that consist of one part that cannot function unless all the elements are completed, can in this respect clarify why cost overruns are higher for fixed link projects. Political-economic explanations consider cost overruns to be the result of the strategic misrepresentation of costs. From this point of view, fixed link projects might be more prestigious, hence decision-makers will do anything in their capacity to get the project realised, e.g. underestimating costs. Lastly, the complexity is usually higher for fixed link projects than for conventional road or rail projects. As a consequence, forecasters, optimistic by nature, will find it more difficult to estimate accurately. In other words, the bias in fixed link projects might be higher. This explanation is based on the psychological notion of optimism bias, the tendency to be overly optimistic. Note that these elucidations only stress that different explanations clarify the differences in average cost overrun between project types; the general belief that cost overruns are mainly the result of political-economic behaviour is not disputed.

A second remarkable finding concerns the improved project performance of rail projects in the new database including 806 projects (from an overrun of 45% to 34%). This can be explained as follows. In the original database the rail projects in the "other geographical areas" had relatively large cost overruns compared to projects in Europe and North America which increased the overall mean. In the new database, the number of projects has increased but only a few of those are projects in the "other geographical areas." The projects in the "other geographical areas" with large overruns hence have less influence with the larger database resulting in a lower average overrun.

**Netherlands versus the Rest of the World**

One of the most remarkable outcomes in the analysis of the Netherlands with the rest of the world is the much smaller average cost overrun for the Dutch projects as compared to that of projects in the rest of the world. Another difference between both databases is the age of the projects; the projects in the Netherlands have been implemented more recently (range of 1991-2010 compared to a range of 1927-2009 for the international projects). If projects that are more recently implemented have lower cost overruns, this could explain the better cost performance in the Netherlands. We therefore tested whether the age influenced the extent of cost overrun. Age could be tested by using either the year of decision to build, the year of construction started or the year of opening as the reference year. The year of opening is used here because the number of projects with information on this variable is the largest (607 projects compared to 338 and 147 projects with information on the year of decision to build and year when construction started respectively). Based on linear regression analyses with cost overrun and year of opening, there is no significant relation between both variables (p-values are 0.173, 0.116, 0.567 and 0.821 for roads, rail, bridges and tunnels). We therefore conclude that age does not influence the cost overrun; age can therefore not explain the difference in average cost overruns between the Netherlands and the rest of the world.

In addition to the average cost overrun, regarding the project types, it turns out that for road and tunnel projects, the Netherlands is not different from the rest of the world. However, for rail and bridge projects Dutch projects perform better, with statistical significance for rail projects although chance cannot be excluded. The statistical insignificance for bridge projects can be explained by the small number of observations for this type of project in the Netherlands.

**Netherlands versus other NW European Countries and Other Geographical Areas**

Comparing the cost performance between the Netherlands, other NW European countries and other geographical areas, it is remarkable that for road projects cost overruns do not vary with geographical location. The average cost overrun varies by only 3% and even the standard deviations are similar between the geographical areas. Also the increase in the number of projects with an additional 370 road projects hardly affected the average cost overrun. The project performance of road projects is relatively stable. This provides opportunities to improve cost estimation procedures for these types of projects.

In addition, the Netherlands has an extraordinary cost performance record for bridge projects, certainly in comparison with projects in the rest of the world which have on average cost overruns up to 4 to 7 times greater. However, the differences are non-significant, probably caused by the small number of Dutch bridges.

Lastly, and probably most remarkable, for rail projects Dutch projects perform significantly better than other NW European countries, which in turn perform better than the rest of the world. Considering the rail projects in more detail, the type of rail could possibly explain the lower average cost overrun. The Dutch rail projects are mostly conventional rail, whereas in the other geographical locations also a considerable number of urban and high-speed rail projects are also included. Although the differences were non-significant, Flyvbjerg et al. (2003b) showed that high-speed rail projects top the list of cost escalation followed by urban rail and conventional rail. Based on the enlarged database, we tested again whether the cost performance of rail projects differed between different types of rail and whether this is different for the Netherlands. The same three types of rail were distinguished: conventional rail, urban rail and high-speed rail (see table 9).

**Table 9 Cost overruns for different rail types in the Netherlands, other NW European countries and other geographical areas**

| Rail Type | The Netherlands | | | Other NW European countries | | | Other geographical areas | | |
|---|---|---|---|---|---|---|---|---|---|
| | N | Mean | SD | N | Mean | SD | N | Mean | SD |
| *Conventional* | 23 | 8.7 | 32.6 | 27 | 20.4 | 37.3 | - | - | - |
| *Urban* | 2 | 10.5 | 23.2 | 23 | 35.5 | 37.0 | 78 | 40.2 | 43.6 |
| *High-speed* | 1 | 55.0 | - | 14 | 26.1 | 25.2 | 4 | 98.8 | 24.0 |
| *Total* | 26 | | | 64 | | | 82 | | |

The representation of Dutch projects for urban rail projects and high-speed rail projects is too sparse to make a meaningful comparison with other regions. The focus is therefore here on conventional rail. Based on a Welch t-test the average cost overrun for conventional rail in the Netherlands is compared with that of other North West European projects. Although the cost overrun on average is smaller for the Netherlands with an average of 9.7% compared to an average of 20.4% for other North West European countries, the difference is not statistically significant (p=0.242). Because of the large variation, the better average performance could be due to chance.

To conclude, almost all projects in the Dutch sample concern conventional rail projects. Comparing these projects with conventional rail projects in the worldwide sample, the difference in average overrun between Dutch projects and other North-West European countries is not statistically significant. Neither is the larger average cost overruns of urban and high-speed rail projects than for conventional rail projects in North-West European (except Dutch) projects statistically significant. The significant difference in cost overrun between Dutch and other North-West European projects is not highly significant, and we cannot conclude whether it is due to a genuine better performance for the Netherlands or a difference between rail types.

## Conclusions and Discussion

Cost overruns are a worldwide phenomenon but a recent study showed that cost overruns in the Netherlands were considerably smaller than in the rest of the world. This paper aimed to establish the extent to which Dutch projects perform significantly better than other geographical areas and whether this differs for project types. Three geographical locations were distinguished: the Netherlands, other NW European countries and other geographical regions. Four project types were considered: road, rail, tunnels and bridges.

The study concludes that worldwide cost overruns differ with project type. Therefore in order to determine whether cost overruns vary with geographical location each project type should be considered separately. For roads and tunnels, the Netherlands performs similar to the rest of the world. For bridge and rail projects, Dutch projects perform considerably better, with statistical significance for the difference in cost overrun for rail projects. Cost overruns of rail projects in the Netherlands are, depending upon the geographical area, 2 to 4 times smaller than in the rest of the world.

These findings have important scientific and policy implications. The study showed with statistical significance, as no other study has previously done, that geographical location matters for project performance, to a varying degree according to project type. It showed that the Netherlands performs better in delivering rail projects than other countries. Since geography matters, there could be other countries with significantly better or worse project performance.

Insight into these countries could provide valuable information about the occurrence of cost overruns. Moreover, as geography matters, there might be other characteristics of countries that can explain the differences in project performance between countries. Countries are different in various aspects, e.g., the decision-making procedures or more generally their system of governance, and this could play a role in project performance as well.

Furthermore, the findings have important policy implications, in particular for the promising new forecasting method called "reference class forecasting (RCF) (Flyvbjerg and Cowi, 2004)." This method achieves accuracy in estimates by basing cost forecasts on actual performance in a reference class of comparable projects thereby bypassing both optimism bias and strategic misrepresentation. Based on the results of this paper, the reference group should be geographically dependent for rail projects. Since the project performance for rail projects differs with geographical location, for future cost forecasts for rail projects the reference group should only contain projects in that specific geographical area. For other types of projects, the reference group can contain projects all over the world. As the geographical location is now taken into account, the overall risk assessment is more detailed and more accurate, thus improving the project management of future projects.

This study was based on data on 806 projects, these were the best obtainable data within our research set. Further efforts to enlarge the database should be made especially for collecting data for projects outside NW Europe. More data are particularly desirable for projects in developing countries. In addition, there are several important issues that need to be addressed in subsequent research. First of all, the cost performance in NW European countries should be examined in more detail and a cross-country comparison would be useful to derive the similarities and differences. Secondly, it should be determined whether other issues can explain the difference in average cost overrun between countries besides to geography, e.g. the decision-making culture or system of governance.